\documentstyle[aps,psfig,amsmath]{revtex}
\newcommand{\be}{\begin{equation}}
\newcommand{\ee}{\end{equation}}
\newcommand{\bea}{\begin{eqnarray}}
\newcommand{\eea}{\end{eqnarray}}
\newcommand{\bef}{\begin{figure}}
\newcommand{\eef}{\end{figure}}
\newcommand{\bm}{\bibitem}
\newcommand{\al}{\alpha}
\newcommand{\bet}{\beta}
\newcommand{\lm}{\lambda}
\newcommand{\sg}{\sigma}
\newcommand{\de}{\delta}
\newcommand{\gf}{\gamma_5}

\newcommand{\gamu}{\gamma_{\mu}}
\newcommand{\gamU}{\gamma^{\mu}}
\newcommand{\om}{\omega}
\newcommand{\rw}{\rightarrow}
\newcommand{\mn}{\mu\nu}
\newcommand{\cl}{{\cal{L}}}
\newcommand{\br}{{\boldsymbol{\rho}}}
\newcommand{\bp}{{\boldsymbol{\pi}}}

\newcommand{\ba}{{\boldsymbol{a}}}
\newcommand{\bv}{{\boldsymbol{v}}}

\newcommand{\F}{F_\pi}
\newcommand{\Fr}{F_\rho}
\newcommand{\dmd}{\partial_\mu}
\newcommand{\dmu}{\partial^\mu}
\newcommand{\dnd}{\partial_\nu}
\newcommand{\dnu}{\partial^\nu}
\newcommand{\emn}{\epsilon_{\mu\nu\lambda\sigma}}
\newcommand{\rt}{\sqrt 2}

\newcommand{\amn}{A_{\mn}}
\newcommand{\fmn}{f_{\mn}}

\newcommand{\omk}{\omega_k}

\newcommand{\bq}{\bar{q}}
\newcommand{\bu}{\bar{u}}

\newcommand{\vk}{\vec{k}}
\newcommand{\vq}{\vec{q}}
\newcommand{\la}{\langle}
\newcommand{\ra}{\rangle}
\newcommand{\ms}{m^2}

\newcommand{\ta}{\tau^a}

\newcommand{\dex}{\Delta_{11}}
\newcommand{\mv}{m_V}
\newcommand{\tmn}{T_{\mn}}
\newcommand{\gmn}{g_{\mn}}
\newcommand{\bmn}{B_{\mn}}
\newcommand{\Lm}{\Lambda}
\newcommand{\Lmn}{\Lm_{\mn}}
\newcommand{\us}{u \!\!\! /}

\newcommand{\tc}{\tau^c}
\begin{document}

\setcounter{page}{1}

\title{Thermal QCD sum rules for mesons}

\author{S. Mallik} 
\address{Saha Institute of Nuclear Physics,
1/AF, Bidhannagar, Kolkata-700064, India} 

\author{Sourav Sarkar} 
\address{Variable Energy Cyclotron Centre, 1/AF, Bidhannagar, Kolkata-700064,
 India}


\maketitle

\begin{abstract} 
A recently proposed scheme is used to saturate the spectral side of the 
QCD sum rules derived from the {\it thermal}, two-point correlation functions of 
the vector and the axial-vector currents. At low temperature, it 
constructs the spectral representation from all the one-loop Feynman 
diagrams for the two-point functions. The old saturation scheme treats   
incorrectly some of these contributions. We end up with the familiar QCD 
sum rules obtained from the difference of the two corresponding {\it vacuum}
correlation functions. The possibility of obtaining new sum rules in other 
media is discussed.

\end{abstract}


\section{Introduction}

In their original work extending the vacuum QCD sum rules~\cite{SVZ}
to those at finite temperature, Bochkarev and Shaposhnikov~\cite{Shap} 
recognised the importance of multiparticle (branch cuts) contributions 
in addition to those of single particles (poles) in constructing the spectral
representation of thermal two-point functions. Thus for the vector current 
correlation function, they included not only the $\rho$-pole with temperature
dependent residue and position, but also the $\pi\pi$ continuum. In the same 
way the spectral representation for the nucleon current correlation function 
would consist of the nucleon pole with temperature dependent parameters and 
the $\pi N$ continuum. Although such a saturation scheme is quite suggestive 
and had been extensively used in the past~\cite{Old,Mallik98}, it lacks a 
theoretical basis, leaving one to suspect if some equally important 
contributions are left out. 

A definitive saturation scheme emerged with the work of Leutwyler and 
Smilga~\cite{Leutwyler90}, who calculated the nucleon current correlation 
function in chiral perturbation theory. Being interested in the shifts of the 
nucleon pole parameters at low temperature, they considered all the one-loop 
Feynman diagrams for the correlation function and evaluated them in the 
vicinity of the nucleon pole. Koike~\cite{Koike} examined the contributions 
of these diagrams in the context of QCD sum rules. He found a new contribution 
arising from the nucleon self-energy diagram to the sum rules, not required in 
the saturation scheme mentioned above.

The one-loop Feynman diagrams for the nucleon correlation function were further
analysed in Ref.~\cite{Mallik02}. In this set of diagrams for the correlation 
function, one has not only diagrams with the (single particle) pole alone and 
the (two particle) branch point alone, but also other (one particle reducible)
diagrams, appearing as product of factors with the pole and the branch cut.
As an example, take the case of a vertex correction diagram having this
product structure. It may be expressed as the sum of the pole term with
constant residue and a remainder, regular at the pole. Clearly to find
the pole parameters one may confine oneself to the pole term alone. But if one
wishes to evaluate the diagram for large space-like momenta -- the region of 
relevance for the QCD sum rules -- both the pole and the remainder become of 
comparable magnitude. It is these remainder terms which are not included 
in the earlier saturation scheme.
 
In this paper we write down the thermal QCD sum rules following from the vector
current and the axial vector current correlation functions, constructing the 
spectral side from the set of all one-loop Feynman diagrams. At low
temperature only the distribution function of the pions is significant in
the heat bath. Thus, of the two particles in the loop, at least one must be a
pion, the other being any one of the strongly interacting particles with 
appropriate quantum numbers in the low mass region. The vertices occurring in 
the diagrams are obtainable from the chiral perturbation theory 
\cite{Gasser84,Weinberg1} as well as from other formulations  
of the effective theory of QCD at low energy \cite{Bando}. But the former 
theory alone is based only on the chiral symmetry of QCD, the others 
assuming one or other unproven symmetries or relations. Thus the chiral 
perturbation theory is singled out to determine the vertices, if we  
claim the sum rules to follow from QCD.

As with the nucleon sum rules \cite{Mallik02}, we subtract out the vacuum sum
rules from the corresponding full sum rules at finite temperature, equating, 
in effect, terms of $O(T^2)$ on both sides. All our calculations are done in 
the chiral symmetry limit, though we keep nonvanishing pion mass in 
intermediate steps.

In Sec. II we collect some results to be used later. In Sec. III 
we analyse the one-loop Feynman diagrams to construct the spectral 
representation for the correlation functions. In Sec. IV we use the known
results of Operator Product Expansion and write the sum rules. Finally our
concluding remarks are contained in Sec.V.  

\section{Preliminaries}
\setcounter{equation}{0}
\renewcommand{\theequation}{2.\arabic{equation}}

Here we review briefly the kinematics of the two point correlation functions 
of the vector and the axial-vector currents. Then we write the interaction 
vertices from the chiral perturbation theory and the results of calculation 
of a prototype loop integral.

\subsection{Kinematics}

Consider the two thermal correlation functions,
\be
T_{\mn}^{ab}=i\int d^4x\, e^{iq\cdot x}\,Tr\,\varrho\,T\,V_\mu^a(x)\,V_\nu^b(0)
~~,
\ee
and
\be
T_{\mn}^{'\,ab}=i\int d^4x\, e^{iq\cdot x}\,Tr\,\varrho\,T\,A_\mu^a(x)\,A_\nu^b(0)
~~,
\ee
of the vector and axial vector currents,
\[V_\mu^a(x)=\bq(x)\gamu\frac{\ta}{^2}q(x),~~~~~~~~~
A_\mu^a(x)=\bq(x)\gamu\gamma_5\frac{\ta}{^2}q(x)~~,\]
generated by the $SU(2)$ flavour symmetry group of the QCD Lagrangian. 
Here $\ta$ are the Pauli matrices and $\varrho=e^{-\beta H}/Tr\,e^{-\beta H}$
is the thermal density matrix of QCD at temperature $T=1/\beta$. Note that 
in the limit of chiral symmetry, in which we shall work, the axial vector 
current is also conserved and the kinematics, in particular, the invariant 
decomposition is the same for both the correlation functions.

The current conservation 
leads to the invariant decomposition
\be
T_{\mn}^{ab}(q)=\delta^{ab}(P_{\mn}T_t+Q_{\mn}T_l),
\ee
where the gauge invariant tensors are chosen as
\[P_{\mn}=-g_{\mn}+\frac{q_\mu q_\nu}{q^2}-\frac{q^2}{\bq^2}\tilde u_\mu\tilde
u_\nu~~,~~~~~~Q_{\mn}=\frac{q^4}{\bq^2}\tilde u_\mu\tilde u_\nu~~,\]
with $\tilde u_\mu=u_\mu-\om q_\mu/q^2$, where $u_\mu$ is the four-velocity
of the medium and $\om$ and $\bq$ are Lorentz invariant scalars,
$\om=u\cdot q$ and $\bq=\sqrt{\om^2-q^2}$, representing the time and space
components of $q_\mu$ in the rest frame of the heat bath $(u_0=1,\,\vec u=0)$.
The invariant amplitudes are functions of the scalar variables, say, $q^2$ 
and $\om$. They can be conveniently extracted from the Feynman diagrams by
forming the scalars,
\be
T_1=g^{\mn}T_{\mn}~~,~~~~T_2=u^\mu u^\nu T_{\mn}~~,
\ee
which are simply related to the invariant amplitudes.

Having cast the kinematics in a Lorentz invariant form, we choose to do
calculations in the rest frame of the heat bath. The kinematic decomposition
(2.3) leads to a constraint on the invariant amplitudes which, in this frame
reads as   
\be
T_t(q_0,\vec q=0)=q_0^2\,T_l(q_0,\vec q=0)~.
\ee
Using this equation, the two sets of amplitudes, $T_{l,t}$ and $T_{1,2}$ can 
be related for $\vq=0$ as
\be
T_l=\frac{1}{\bq^2}T_2~~,~~~~~T_t=-\frac{1}{3}T_1~~ .
\ee
Note also the symmetry of the imaginary parts of the amplitudes,
\be 
Im T_{l,t}(-q_0,\vq=0)=Im T_{l,t}(q_0,\vq=0).
\ee

In the real time thermal field theory, we are going to use here,
each of the above amplitudes stands for a $2\times 2$ matrix, whose
components depend on a single analytic function~\cite{TFT}. 
This function, in turn,
is determined completely by the 11-component function itself: their
real parts are equal and the imaginary part of the former equals
$\pi^{-1}tanh(\bet q_0/2)$ times that of the latter. (The factor $\pi^{-1}$
is inserted for convenience.) To avoid further symbols, we henceforth
redefine $T$ to denote this analytic function. Its spectral 
representation at fixed $\vq$ is given by,
\be
T_{l,t} (q_0^2, \vq)= \int_0^\infty
\frac{dq_0^{\prime 2}\,Im\,T_{l,t}(q_0^\prime,\vq)}{q_0^{\prime
2}-q_0^2-i\epsilon}.
\ee

\subsection{Dynamics}

As we pointed out already, we must describe interaction among particles
according to the chiral perturbation theory.
The calculation of the two point functions in this theory is most conveniently
carried out in the external field method \cite{Gasser84}, where 
the original QCD Lagrangian is extended by introducing external fields 
$v_\mu^a(x)$ and
$a_\mu^a(x)$ coupled to the currents $V_\mu^a(x)$ and
$A_\mu^a(x)$,
\[{\cal L}_{QCD}\rw{\cal L}_{QCD}+v_\mu^a(x)V^\mu_a(x)+a_\mu^a(x)A^\mu_a(x).\] 
The resulting interaction vertices are obtained in Ref.\cite{Ecker89,Ecker89a} 
for the symmetry group $SU(3)_R\times SU(3)_L$.

At low temperature the pions dominate the heat bath. We thus consider the
reduced symmetry $SU(2)_R\times SU(2)_L$ and write down the chiral
couplings of pions with the external fields and the observed particles, to be
encountered in the one-loop diagrams \cite{Preprint}. We leave out the 
interaction among the pions themselves, as it would give rise to corrections 
of order higher than $T^2$. Their interaction with external fields are
given by 
\be
\cl_{int}(\pi)=\cl_v(\pi)+\cl_a(\pi)
\ee
where
\bea
\cl_v(\pi)&=& \bv_\mu\cdot\bp\times\dmu\bp+
\frac{1}{2}\left(\bp\cdot\bp \,\bv_\mu\cdot\bv^\mu-
\bv_\mu\cdot\bp\, \bv^\mu\cdot\bp\right)\nonumber\\
\cl_a(\pi)&=&\frac{1}{2}\F^2\ba_\mu\cdot\ba^\mu-\F\ba_\mu\cdot\dmu\bp
-\frac{1}{2}\left(\bp\cdot\bp\,\ba_\mu\cdot\ba^\mu-
\ba_\mu\cdot\bp \,\ba^\mu\cdot\bp\right)\nonumber\\
&&+ \frac{1}{2\F}\left(\bp\cdot\bp\,\dmu\bp\cdot\ba_\mu-
\bp\cdot\dmu\bp\,\bp\cdot \ba_\mu \right)~,
\eea
the letters in bold face denoting isospin vectors.
The third and the fourth terms in $\cl_a(\pi)$ represent corrections
respectively to the first and the second term, when two of the pion fields
without derivatives are contracted in them. Such a contraction can be
worked out from the thermal pion propagator
\be
\dex^{(\pi)} (k)= \frac{i}{k^2-m_\pi^2} +2\pi\de (k^2-m_\pi^2)n(k),
\ee
where $n(k_0) =(e^{\bet |k_0|} -1)^{-1} $ is the Bose distribution function. 
Then the thermal part of the contraction gives
\be
Tr\,\varrho\,T\pi^a(x)\pi^b(x) |_{11} \rw
\de^{ab}\int\,\frac{d^4k}{(2\pi)^3}n(k_0)\delta(k^2-m_\pi^
2)=\de^{ab}\frac{T^2}{12},
\ee
in the chiral limit. We thus obtain the effective form of $\cl_a(\pi)$ as
\be
\cl_a(\pi)=\frac{\F^2}{2}\left(1-\frac{T^2}{6\F^2}\right)\ba_\mu\cdot\ba^\mu
-\F\left(1-\frac{T^2}{12\F^2}\right)\ba_\mu\cdot\dmu\bp~.
\ee

Next we write the couplings of the isotriplets
$[\rho (770), a_1 (1230)]$ and isosinglets $[\om (782), f_1 (1282)]$
of vector $(1^{- -})$ and axial vector $(1^{+ +})$ mesons respectively.
Although we take their fields to transform according to $SU(2)$ in
constructing their interactions, we take the physical (zero temperature)
masses of the multiplets of each of the $SU(3)$ octets to be degenerate. 
Then the coupling linear in the vector meson fields are given by
\bea
\cl (V)&=&\frac{F_{\rho}}{m_V}\left\{\left(1-\frac{T^2}{12\F^2}\right)\dmu\bv^\nu\cdot
(\dmd\br_\nu-\dnd\br_\mu)
+ \frac{1}{\F}\dmu\ba^\nu\cdot(\dmd\br_\nu-\dnd\br_\mu)\times{\bp}
\right\}\nonumber \\
& &-\frac{2G_{\rho}}{m_V\F^2}\dmd\br_\nu\cdot\dmu\bp\times\dnu\bp
-\frac{\rt H_\om}{m_V\F}\emn \om^{\mu}\dnu\bp\cdot
\partial^\lm\bv^\sg ,
\eea
while those linear in the axial vector meson fields are
\bea
\cl (A)&=&-\frac{F_{a_1}}{m_A}\left\{\left(1-\frac{T^2}{12\F^2}\right)
\dmu\ba^\nu\cdot(\dmd\ba_{1\nu}-\dnd\ba_{1\mu})
+ \frac{1}{\F}\dmu\bv^\nu\cdot(\dmd\ba_{1\nu}-\dnd\ba_{1\mu})\times{\bp}
\right\}\nonumber \\
& & +\frac{\rt H_{f_1}}{ m_A\F}\emn f_1^{\mu}\dnu\bp\cdot
\partial^\lm\ba^\sg.
\eea
Finally the quadratic couplings of the triplets with the singlets and
between themselves are given by
\be
\cl (V,A)=-2\emn\left( \frac{g_1}{\F}\dmu\om^\nu
\br^\lm\cdot\partial^\sg\bp +\frac{g_2}{\F}
\dmu f_1^\nu  \ba_1^\lm\cdot\partial^\sg\bp\right) +
\frac{g_3}{\F}
\dmu\br^\nu\cdot(\ba_{1\,\mu}\times\dnd\bp-\ba_{1\,\nu}\times\dmd\bp).
\ee
The coupling constants in the above interaction terms can be determined from
the decay rates of the particles \cite{Ecker89a,Preprint}. The propagator for 
the heavy meson can be taken to be the one in vacuum.
 
There also appear two other spin-one mesons $h_1(1170)$ and $b_1(1235)$
having quantum numbers $1^{+-}$~\cite{PDG}. But $P$ and $C$ invariance 
forbids their appearance in our diagrams \cite{Ecker89}. The pseudoscalar 
$SU(2)$ singlets $\eta (547)$ and $\eta^{\prime} (985)$ also have no couplings 
relevant for us. The scalars $\sigma (400-1200)$ and $a_0 (980)$ do have 
vertices for our diagrams, but they do not contribute to $O(T^2)$, as do many 
of the vertices written above. We shall omit the vertices with the scalars 
altogether from our discussions to follow.

With the above interaction vertices we can write the Feynman amplitudes 
for the correlation functions. A loop in the diagrams makes a
contribution of the form
\be
F_{\mn} (q)=ic\int \frac{d^4k}{(2\pi)^4} \fmn (q,k) \dex^{(\pi)} (k)
 \dex^{(X)} (q-k),
\ee
where the particle $X$ can be a heavy (spin one) meson or the pion itself.  
The tensor $\fmn$ is given by the interaction vertices and any tensor
structure in $\dex^{(X)}$. Out of the interaction vertices we have isolated
the coupling constants in $c$. Being gauge invariant, $F_{\mn}$ allows one 
to construct the invariant amplitudes $F_{l,t}$ in the same way as we did 
above for $T_{\mn}$. 

As we shall see below, the one-particle irreducible diagrams consist only of 
loop integrals like (2.17), while the reducible ones are given by such
integrals multiplied with a heavy particle
pole (of first or second order). In the former case we may evaluate directly
the leading contribution of the integral and then Borel transform. In the
latter case it is more systematic to use the exact spectral representation
of the loop integral. We then Borel transform the resulting complete 
amplitude and extract its leading term. 

Until now we have considered full amplitudes, including the corresponding
vacuum amplitudes. Our thermal sum rules need only their temperature dependent
parts. So we evaluate only these (convergent) parts of the loop integrals,
but continue to denote them by the same symbols as for the full amplitudes.

Consider first the case where $X$ is a heavy particle of mass $m_H$. Let the
full amplitude be given by just the loop integral (2.17). Going over to the
invariant amplitudes, its thermal part is
as 
\be
 F_{l,t} (q) =-c\int\frac{d^4k}{(2\pi)^3} \frac{\de (k^2-m_\pi^2)
n(k)}{(q-k)^2 -m_H^2} f_{l,t}(q,k),
\ee
whose leading behaviour for large space-like momenta with $\vq=0$ 
$(q_0^2=E^2=-Q^2<0)$ is given by $(\om_k=\sqrt{\vk^2 +m_\pi^2})$
\be
F_{l,t} (Q^2,\vq =0) \rw \frac{2c}{Q^2 +m_H^2} \int\frac{d^3k \,
n(k)}{(2\pi)^3 2\om_k} f_{l,t}(Q,|\vk|).
\ee
It is clear that only if $f_{l,t}(Q,|\vk|)$ is constant in $\vk$, is the
integral of order $T^2$; otherwise, it is of higher order. 

To obtain the spectral representation of the loop integral for $\vq =0$, 
we note that the
cuts in $E^2$ plane in this case are given by $0<E^2<(m_H -m_\pi)^2$ and
$E^2 >(m_H + m_\pi)^2$. The imaginary parts across both these cuts are given
by \cite{Mallik02}
\be
Im F_{l,t}(E)= \frac{\sqrt{\om^2-m_\pi^2}}{8\pi^2 E}n(\om)\,f_{l,t}(E,\om)
\ee
where $\om=(E^2-m_H^2+m_\pi^2)/2E$. We now have the spectral representation
for $F_{l,t}$ given by an equation like (2.8), the integration running over 
the two cuts stated above.

If the particle $X$ is also a pion, the leading term of the loop integral
may again be found in the same way as we did above for the heavy particle.
However, to find the spectral representation for $\vq =0$, one must begin
with $\vq \neq 0$ and then go to the limit. This evaluation is done in the
Appendix.

\section{Spectral representation}
\setcounter{equation}{0}
\renewcommand{\theequation}{3.\arabic{equation}}

Here we analyse the different one-loop Feynman diagrams for the 
correlation functions and extract the leading thermal contributions (to order 
$T^2$) to the spectral
side of the sum rules. These diagrams are the same as those 
considered in \cite{Preprint} to find the shifts in the pole parameters of 
the $\rho$ and
$a_1$ mesons. The difference lies in the evaluation of the diagrams: While
we evaluated them earlier in the neighbourhood of the respective poles, now we
have to do so at large space-like momenta.

The diagrams for the correlation functions can be grouped into three types,
namely those with intermediate states, vertex corrections and 
self-energies.
To write the Feynman amplitudes we anticipate the 
following gauge invariant tensors,
\bea
\amn(q)&=&-\gmn+{q_\mu q_\nu}/{q^2}=P_{\mn} +Q_{\mn}/q^2 ,\nonumber\\
\bmn(q,k)&=&q^2 k_\mu k_\nu-q\cdot k(q_\mu k_\nu+k_\mu q_\nu)
+(q\cdot k)^2\gmn ,\nonumber\\
C_{\mn}(q,k)&=&q^4 k_\mu k_\nu-q^2(q\cdot k)(q_\mu k_\nu+k_\mu q_\nu)
+(q\cdot k)^2q_\mu q_\nu .
\eea
We now consider the diagrams separately for the
correlation functions of the vector and the axial vector currents. 
Although we need only the $T$-dependent parts of the diagrams to
order $T^2$, we shall write the pole amplitudes in full. 

\subsection{Vector current}

Here the pole and the one-loop diagrams are shown in Figs. 1-4.
 
\bef
\centerline{\psfig{figure=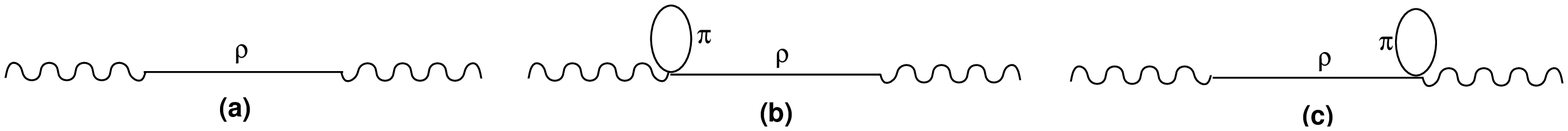,height=1.0cm,width=12cm}}
\caption{ $\rho$ pole and constant vertex correction}
\eef

\bef
\centerline{\psfig{figure=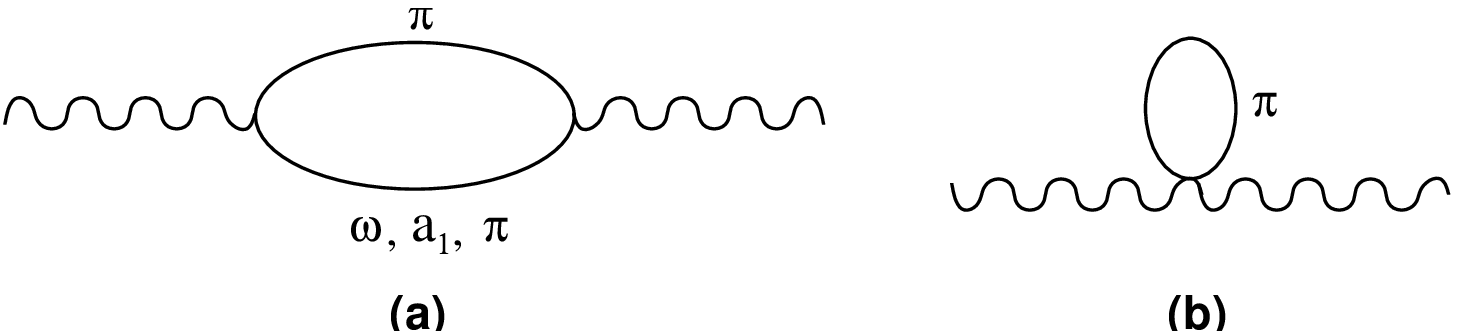,height=1.5cm,width=6cm}}
\caption{Intermediate state diagrams}
\eef

\bef
\centerline{\psfig{figure=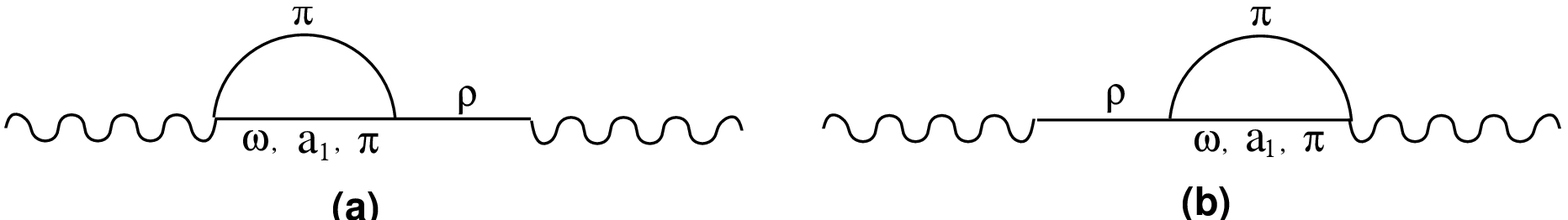,height=1.0cm,width=7cm}}
\caption{Vertex correction diagrams}
\eef

\bef
\centerline{\psfig{figure=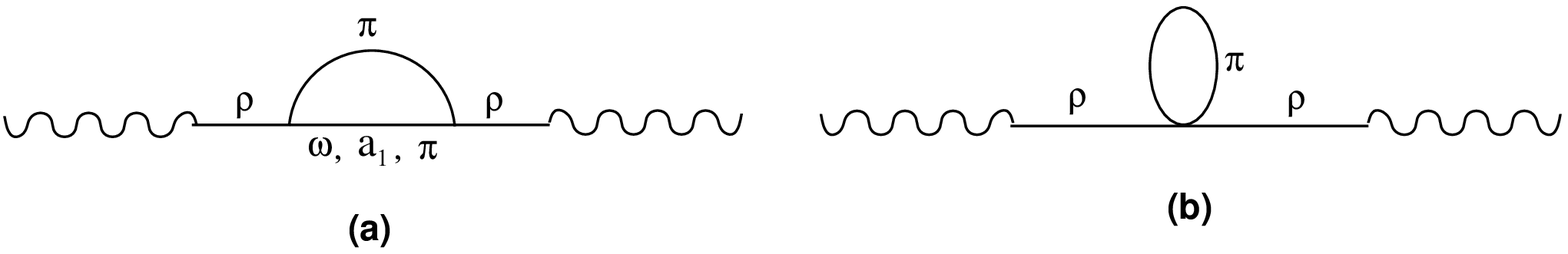,height=1.5cm,width=7cm}}
\caption{Self-energy diagrams}
\eef

The $\rho$ pole amplitude with its constant vertex corrections (Fig. 1)
is given by
\be
T_{\mn}^{(\rho)}(q)=-\left(\frac{\Fr}{\mv}\right)^2
\left(1-\frac{T^2}{6\F^2}\right)\frac{q^4}{q^2-\mv^2}\amn
\ee 

The amplitudes for the intermediate state diagrams (Fig. 2) are generally
of the form of Eq.(2.17). In the case of the $\pi\pi$ intermediate state, one
has to include also the tadpole diagram of Fig. 2(b) giving
\be
\tmn^{(\pi\pi)} (q)=i\int \frac{d^4k}{(2\pi)^4} \left\{(2k-q)_\mu (2k-q)_\nu
\dex^{(\pi)} (k) \dex^{(\pi)} (q-k)-2i\gmn\dex^{(\pi)} (k)\right\}, 
\ee
whose thermal part is
\be
\tmn^{(\pi\pi)} (q)= \amn\,(-2q^2)\int\frac{d^4k}{(2\pi)^3}
\frac{\de (k^2-\ms_\pi)n(k)}{(q-k)^2-\ms_\pi} \rw -\frac{T^2}{6} \amn
\ee
at large space-like momenta.
For the $\pi a_1$ intermediate state, the amplitude is given by Eq. (2.17)
with
\[ c=2(F_{a_1}/m_A\F)^2,~~~ \fmn=q^2(q^2-2q\cdot k)\amn-\bmn,\]
getting 
\[f_{l,t}=E^2(1,\,E^2)~.\]
Then Eq.~(2.19) gives 
\be
T_{l,t}^{(\pi a_1)}\rw
\left(\frac{F_{a_1}}{m_A}\right)^2 \ \frac{(-Q^2,\,Q^4)}{Q^2+m_A^2}
\cdot\frac{T^2}{6\F^2}
\ee
The (Borel transforms of the) amplitudes given by Eq.~(3.2,3.4,3.5) constitute
all of the contributions to the spectral side of the sum rules. In the rest of 
this subsection we verify that none of the remaining diagrams with 
$\pi\om$ intermediate state, vertex corrections
and self-energies contribute to order $T^2$ to the sum rules. Clearly it 
suffices to
show this behaviour for any one of the invariant amplitudes, say $T_l$ and
we omit its subscript in the following.

For the $\pi\om$ intermediate state, the amplitude is again given by
Eq.~(2.17) with
\be
c=-2(H_\om/\mv\F)^2,~~~ \fmn=q^2k^2\amn+\bmn,~~~ f=-2|\vk|^2/3~,
\ee
where and below $f$ stands for $f_l$.
Then Eq.~(2.21) shows immediately that this amplitude is of order $T^4$.

Considering next the vertex corrections of Fig.~3, 
each of the amplitudes is of the form
\be
\frac{F_\rho}{\mv}\frac{q^2}{q^2-\mv^2} \Lm_{\mn} ~~,
\ee
where $\Lm_{\mn}$ is again a one-loop integral of the form of Eq.~(2.17). 
Consider now
\be 
\frac{E^2}{E^2-\mv^2}\Lm (E^2),
\ee 
where the invariant amplitude $\Lm (E^2)$ satisfies the spectral representation
\[ \Lm (E^2)=\int\frac{Im\Lm (E^{\prime})\,dE^{\prime 2}}
{E^{\prime 2}-E^2}~,\] 
which may be used to write the expression (3.8) as
\be
\frac{\mv^2\Lm(\mv^2)}{E^2-\mv^2}+\int\frac{dE^{\prime 2}E^{\prime 2}
Im\Lm (E^{\prime})}{(E^{\prime 2}-\mv^2)(E^{\prime 2}-E^2)}~,
\ee
separating the pole term from the one regular in the vicinity of the pole. 
On the other hand, if we go to large space-like $E^2=-Q^2$ and take the Borel
transform, it becomes
\be
\frac{e^{-\mv^2/M^2}}{M^2}\int\frac{dE^2\,Im\Lm (E)}{(E^2-\mv^2)}
\left(-\mv^2+E^2e^{-(E^2-\mv^2)/M^2}\right)~,
\ee
where $M^2$ is the Borel variable replacing $Q^2$.

Eqs.~(3.9) and (3.10) allow us to compare how the same amplitude behaves 
near the pole and at large space-like momenta in the form of the Borel transform. 
Consider the vertex correction from the $\pi\om$ loop for which
$c=-(4\sqrt{2}H_\om g_1/\mv\F^2)$ and the expressions for $\fmn$ and $f$ are 
identical to those in Eqs. (3.6). Then we see that in Eq.~(3.9)
$\Lm(\mv^2)\sim O(T^4)$ and the second term is indeed finite at $E^2=\mv^2$
(and of order $T^2$). Next consider the behaviour of the Borel transform
(3.10). Because the branch point of the function $\Lm (E^2)$ also starts at 
$E^2=\mv^2$, its leading term is given by 
\be
\left(1-\frac{\mv^2}{M^2}\right)\frac{e^{-\mv^2/M^2}}{M^2}\int 
dE^2\,Im\Lm (E)~,
\ee
which is of order $T^4$. Thus to $O(T^2)$, the vertex correction diagram 
with the $\pi\om$ loop contributes neither to the pole parameters
nor to the Borel transform to $O(T^2)$. Notice, however, that while near 
the pole the second term in Eq.~(3.9) can be ignored, for large 
space-like momenta both the terms assume equal importance. 

Consider next the $\pi a_1$ loop with
\[c=-(4F_{a_1}g_3/m_A\F^2),~~~ \fmn=q^2q\cdot k\amn+\bmn ,~~~
f=E|\vk|~.\]
Again we see that at $E^2=\mv^2$, the residue $\Lm(\mv^2)\sim O(T^4)$.
Here the branch point of $\Lm (E^2)$is is at $E^2=m_A^2$. So the Borel transform
is clearly of order $T^4$. As shown in the Appendix, the $\pi\pi$ loop contribution
is also at least of order $T^4$ to the pole residue and Borel transform.

Finally the self-energy diagrams of Fig.~4 can be analysed in a similar way. 
Each of the diagrams contributes an amplitude of the form
\be
-\left(\frac{\Fr}{\mv}\right)^2\cdot \frac{q^4}{(q^2-\mv^2)^2}\Pi_{\mn}(q)~.
\ee
The self energy $\Pi_{\mn}$ is again of the form of Eq.~(2.17). Using the  
spectral representation for its invariant amplitudes, we get 
\bea
&&\frac{E^4}{(E^2-\mv^2)^2}\Pi (E)\nonumber\\
&&=\frac{\mv^4}{(E^2-\mv^2)^2}\int\frac{dE^{\prime 2}
Im\Pi (E^{\prime})}{E^{\prime 2}-\mv^2}
+\frac{\mv^2}{E^2-\mv^2}\int\frac{dE^{\prime 2}(2E^{\prime 2}-\mv^2)
Im\Pi (E^{\prime})}{(E^{\prime 2}-\mv^2)^2}\nonumber\\
&&+\int\frac{dE^{\prime 2}E^{\prime 4}
Im\Pi (E^{\prime})}{(E^{\prime 2}-\mv^2)^2(E^{\prime 2}-E^2)}~,
\eea
which is the
appropriate expression to study the neighbourhood of the meson pole. If we
now go to space-like momenta and take the Borel transform, we get
\be
\frac{1}{M^2}\int dE^2\,Im\Pi (E)\left[\left\{1-\left(\frac{E^2}
{E^2-\mv^2}\right)^2+\frac{\mv^2}{M^2}\frac{\mv^2}{E^2-\mv^2}
\right\}e^{-\mv^2/M^2}+\left(\frac{E^2}{E^2-\mv^2}\right)^2
e^{-E^2/M^2}\right]
\ee

Consider first the $\pi\om$ self-energy loop, whose evaluation reveals 
the inadequacy
of the old saturation scheme. For this loop we have
$c=4(g_1/\F)^2$ and the expressions for $\fmn$ and $f$ are again given by
Eqs. (3.6).
Near the meson pole, the three integrals in (3.13) are all finite and are of
order $T^4 ,T^2$ and $T^2$ respectively. Thus the pole position does not
shift, but the residue does to order $T^2$ \cite{Preprint}. On the other hand
the Borel transform (3.14), which simplifies to leading order as 
\[
e^{-\mv^2/M^2}\int dE^2\,
\left(1-\frac{E^2+\mv^2}{M^2}+\frac{E^4}{2M^4}\right)\,Im\Pi (E),\]
is clearly of order $T^4$. To summarise, the self-energy diagram with the
$\pi\om$ loop gives a correction to the pole residue to order $T^2$, but the
Borel transform of the full amplitude has no contribution to this order. By 
contrast, the old saturation scheme would retain the correction to
the residue also in the Borel transform. It amounts to the neglect of the
third, regular term in Eq. (3.13), which is of course justified near the
pole, but not for large space-like momenta, where the
pole and the regular terms are not only of comparable magnitudes, but
actually cancel each other in the leading order.   
 
The $\pi a_1$ self-energy loop has
\[c=2(g_3/\F)^2,~~~ \fmn=\bmn-C_{\mn}/m_A^2,~~~f=-|\vk|^2(2+E^2/m_A^2)/3~.\]
We see that this amplitude contributes neither to the pole position
nor to the Borel transform to $O(T^2)$. The same is the case with the $\pi\pi$
loop as shown in the Appendix.

\subsection{Axial-vector current}

Here the Feynman diagrams are shown in Figs. 5-9. The new feature here is 
the existence of the pion pole in addition to the one
for the axial vector meson $a_1$.

\bef
\centerline{\psfig{figure=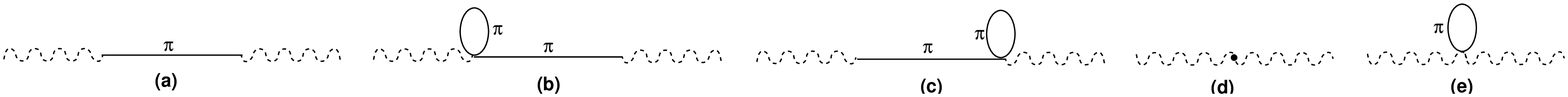,height=1.0cm,width=18cm}}
\caption{ $\pi$ pole and constant vertex corrections}
\eef

\bef
\centerline{\psfig{figure=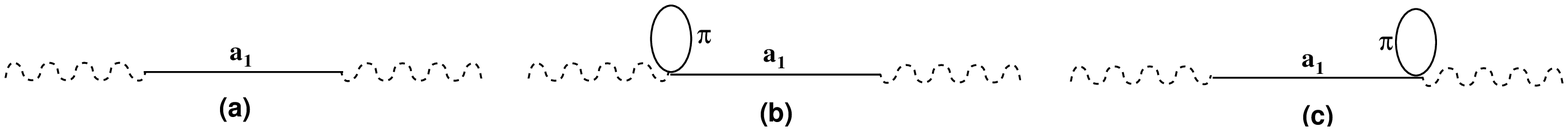,height=1.0cm,width=11cm}}
\caption{ $a_1$ pole and constant vertex corrections}
\eef

\bef
\centerline{\psfig{figure=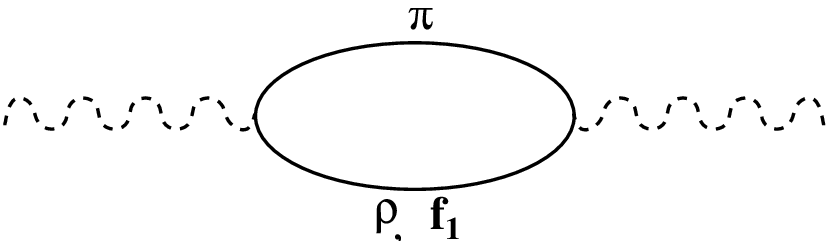,height=1.0cm,width=3.5cm}}
\caption{Intermediate state diagrams}
\eef

\bef
\centerline{\psfig{figure=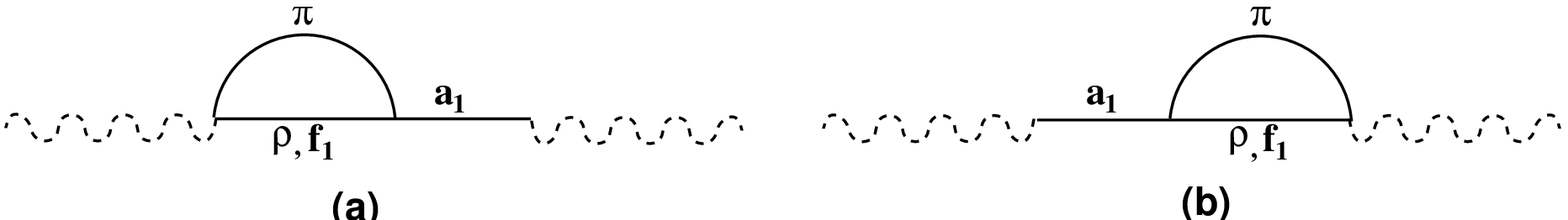,height=1.0cm,width=7cm}}
\caption{Vertex correction diagrams}
\eef

\bef
\centerline{\psfig{figure=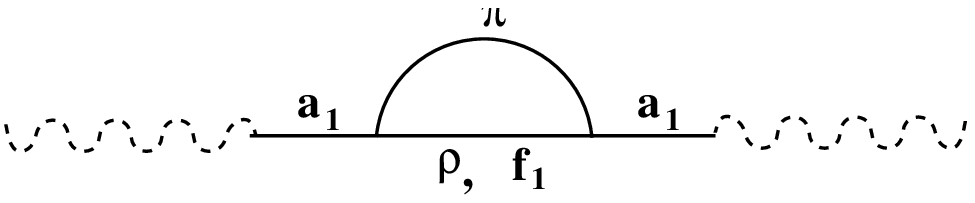,height=1.5cm,width=3.5cm}}
\caption{Self-energy diagrams}
\eef

The $\pi$-pole with vertex corrections of Fig.~5 is given by
\be
T_{\mn}^{\prime (\pi)}=-\F^2\left(1-\frac{T^2}{6\F^2}\right)\amn
\ee
in the chiral limit. The $a_1$-pole with vertex corrections of Fig.~6 is given 
by
\be
T_{\mn}^{\prime (a_1)}=-\left(\frac{F_{a_1}}{m_A}\right)^2
\left(1-\frac{T^2}{6\F^2}\right)\frac{q^4}{q^2-m_A^2}\amn~.
\ee
The contribution of the $\pi\rho$ intermediate state of Fig.~7 to the invariant
amplitudes $T_{l,t}$ are given by
\be
T_{l,t}^{\prime
(\pi\rho)}=\left(\frac{F_{\rho}}{m_V}\right)^2 \frac{T^2}{6\F^2}
\frac{(-Q^2,\,Q^4)}{Q^2+m_V^2}.
\ee
As in the case of the vector current correlation function, one can show that
none of the remaining diagrams contribute to order $T^2$ at large space-like
momenta.

\section{Sum Rules}
\setcounter{equation}{0}
\renewcommand{\theequation}{4.\arabic{equation}}

We now turn to the operator side of the sum rules. Up to dimension six, the
Lorentz scalar operators appearing in the short distance expansion of the 
product of two currents are $\bf 1$, $m_q \bar q q$, $G_{\mn}G^{\mn}$ and 
two four-quark operators. Of the latter two, one arises from the Taylor 
expansion of the
quark bilocal operators arising in the Wick expansion of the currents. 
This piece turns out to be isospin scalar. 
The other four-quark operator arises from the Feynman diagram to second order 
in QCD perturbation expansion, where the large external momentum flows 
through the internal gluon line.
 
In the medium there are additional operators, which are Lorentz non-scalars in 
the rest frame of the heat bath. They can, of course, be written as Lorentz 
scalars in a general frame by using the four-velocity vector $u_\mu$. 
They first appear as dimension four, which are $\bq \us q$, 
$u^\mu u^\nu\Theta_{\mn}^f$ and
$u^\mu u^\nu\Theta_{\mn}^g$, where $\Theta_{\mn}^{f,g}$ 
are the energy momentum tensors of the quark and gluon respectively. 
More such operators arise at dimension five and six, but generally contain 
derivatives.

In the process of subtracting out the vacuum sum rule to extract terms of 
order $T^2$, the unit operator drops out. Also drops out $m_q\bq q$ in the 
chiral limit. The isospin scalar four-quark operator cannot have a
temperature dependence in the chiral limit. 
The operator $\bq \us q$ does not contribute, as we do not 
have non-zero chemical potential for the fermions. The thermal expectation 
value of $G_{\mn}G^{\mn}$, $\Theta_{\mn}^{f}$ and $\Theta_{\mn}^{g}$ are all 
of order $T^4$. Also the expectation values of other operators with derivative(s) 
are of higher order than $T^2$, as each derivative gives rise to an
additional power in pion momentum.

Thus we are left with only the isospin non-scalar four quark operator as 
the relevant one for our sum rules. Their Wilson coefficients are 
known \cite{SVZ},
\bea
i\int d^4x\, e^{iq\cdot x}\,\la T\,V_\mu^a(x)\,V_\nu^b(0)\ra&\rw&
\de^{ab}\left(
-\gmn+\frac{q_\mu q_\nu}{q^2}\right)\frac{8\pi}{3Q^4}\la O_A\ra\nonumber\\
i\int d^4x\, e^{iq\cdot x}\,\la T\,A_\mu^a(x)\,A_\nu^b(0)\ra&\rw&
\de^{ab}\left(
-\gmn+\frac{q_\mu q_\nu}{q^2}\right)\frac{8\pi}{3Q^4}\la O_V\ra
\eea
where
\begin{eqnarray*}
O_V=\al_s\bq\gamu\frac{\tc}{^2}\frac{t^i}{^2}q\,\bq\gamU\frac{\tc}{^2}
\frac{t^i}{^2}q\nonumber\\
O_A=\al_s\bq\gamu\gf\frac{\tc}{^2}\frac{t^i}{^2}q\,\bq\gamU\gf\frac{\tc}{^2}
\frac{t^i}{^2}q\nonumber\\
\end{eqnarray*}
and $\al_s=g_s^2/4\pi$ is the strong interaction fine structure constant.

The thermal average of any operator may be expanded as 
\be
\la O\ra=\la 0|O|0\ra+
\sum_a\int\frac{d^3k\, n(k)}{(2\pi)^32\omk}\la\pi^a(k)|O|\pi^a(k)\ra +\cdots,
\ee
where the sum is over the isospin indices of the pion. The pion matrix
elements of $O_{V,A}$ are easily worked out using PCAC and Current Algebra 
to get,
\[\la O_{V,A}\ra=\la 0| O_{V,A}|0\ra\mp\frac{T^2}
{6\F^2}\la 0| O_V-O_A|0\ra\]

It is now simple to write the sum rules for a correlation function by equating 
the Borel transform of the spectral representation to the operator product
expansion. It turns out that both the vector and the axial vector correlation 
functions give rise to the same sum rules \cite{Comment},
\bea
-\Fr^2e^{-\mv^2/M^2}+ F_{a_1}^2e^{-m_A^2/M^2}+\F^2
&=&-\frac{4\pi}{3M^4}\la 0| O_V-O_A |0\ra\nonumber\\
-\mv^2\Fr^2e^{-\mv^2/M^2}+m_A^2F_{a_1}^2e^{-m_A^2/M^2}
&=&\frac{8\pi}{3M^2}\la 0|O_V-O_A |0\ra
\eea

It is interesting to observe that these sum rules are nothing but the vacuum 
sum rules derived from the 
difference of the correlation functions of the vector and 
the axial-vector currents, 
whose spectral side consists of the pole terms from $\pi$, $\rho$ and 
$a_1$ exchanges. Also as $M^2$ tends to infinity, we recover the well known 
sum rules
\bea
\Fr^2-F_{a_1}^2-\F^2=0\nonumber\\
\mv^2\Fr^2-m_A^2F_{a_1}^2=0
\eea
originally derived by Weinberg \cite{Weinberg2} as superconvergence sum rules 
for the difference of the spectral functions.

\section{Concluding remarks}

In this work we have obtained the sum rules following from the two point
functions of the vector current and the axial vector current at finite
temperature. For a reliable estimate, we calculate their spectral sides
using the chiral perturbation theory. The resulting Feynman diagrams (to one
loop) can be readily evaluated for their leading thermal contributions. 

Though there is a multitude of diagrams to begin with, only a few of these
actually contribute to order $T^2$. This is due to the presence of
derivatives on the pion fields in the interaction vertices derived from the
chiral perturbation theory. Among the contributing ones, there is a
one-particle reducible, self-energy diagram, whose contribution at space-like 
momenta is zero to order $T^2$. But the old saturation scheme would derive
from this diagram a shift to this order in the residue of the meson pole.

The older strategy of determining the temperature dependence of the pole
parameters from such sum rules is not relevant anymore, as we have already
included the diagrams responsible for this dependence in our calculation of
the spectral functions. Our evaluation reproduces the sum rules which follow 
from the vacuum correlation functions themselves. 

The present saturation scheme has been used also for the nucleon correlation
function at finite temperature, again reproducing the results obtainable 
from the corresponding vacuum correlation function \cite{Koike,Mallik02}. 
Thus we now have enough confirmation of the correctness of the QCD 
sum rules in a medium.

Can we get {\it new} results from the sum rules in a medium? 
It would seem that we are merely reproducing the old results in a more
complicated way: What one could have from single particle exchange diagrams
in vacuum, are now derived from one loop diagrams at
finite temperature. But we do not believe the situation to be so. Indeed, 
having been assured of the correct procedure to saturate the spectral side, we
may apply the sum rule technique to other media, like the nuclear medium
by introducing the nucleon chemical potential. There are no reliable estimates
of higher dimension operators like the four-quark operators in such media.
It is these quantities which should be readily available from such sum rules.

\section*{Appendix}
\setcounter{equation}{0}
\renewcommand{\theequation}{A.\arabic{equation}}

Some care is necessary in constructing the spectral representation for the
$\pi\pi$ loop for $\vq=0$. The loop integral gives rise to cuts in the $q_0^2$
 plane for arbitrary $\vq$ extending over $0<q_0^2<|\vq|^2$ (short cut) 
and $q_0^2>4(m_\pi^2+|\vq|^2)$ (unitary cut). As $\vq\rw 0$ the short cut 
shrinks to a point. But the spectral function can be singular in this limit, 
so that the part of the spectral representation due to this cut may be 
non-vanishing.

Consider first the $\pi\pi$ loop in the vertex diagram. 
It gives an amplitude of the form of Eq. (3.7), where
\be
\Lmn (q)=ic\int \frac{d^4k}{(2\pi)^4} \fmn (q,k) \dex^{(\pi)} (k)
 \dex^{(\pi)} (q-k),
\ee
with \[ c=(4G_{\rho}\Fr/\mv^2\F^2)~, 
~~\fmn=(q_\mu-2k_\mu)(q^2k_\nu-q\cdot kq_\nu) ~.\]
As in Sec. II the imaginary part of the amplitude may again be obtained 
directly by integrating over $k_0$. But, unlike the case of loops with one 
pion and one heavy particle, the angular integration in the resulting integral 
over $\vk$ gives rise to a $\theta$-function constraining the limits of $|\vk|$. 
The imaginary part of the invariant amplitudes may then be worked out to give
\be
\left(\begin{array}{c}Im\Lm_l\\ Im\Lm_t\end{array}\right)=
\frac{1}{8(2\pi)^2}\int_{-1}^{+1}dx\left(\begin{array}{c}
-q^2x^2/2\\-q^4/6\end{array}\right)\left(1+\frac{2}{e^{\bet(|\vq|x+q_0)/2}-1}
\right)~,~~q^2>0~,
\ee
and
\be
\left(\begin{array}{c}Im\Lm_l\\ Im\Lm_t\end{array}\right)=
\frac{1}{8(2\pi)^2}\int_{1}^{\infty}dx\left(\begin{array}{c}
-q^2x^2/2\\-q^4/6\end{array}\right)\left(\frac{1}{e^{\bet(|\vq|x-q_0)/2}-1}
-\frac{1}{e^{\bet(|\vq|x+q_0)/2}-1}\right)~,~~q^2<0~.
\ee

On the unitary cut the $x$-integrals are thus finite as $\vq \rw 0$ and
because of $q^2(q^4)$ in the integrands, their Borel transforms will have no
contributions to $O(T^2)$. But on the short cut the $x$-integrals diverge as 
$\vq \rw 0$. Let us then consider the full Borel transformed amplitude with
$\vq \neq 0$,
\be
\left(\begin{array}{c}\Lm_l\\ \Lm_t\end{array}\right)=
\frac{1}{M^2}\int_0^{|\vq|^2}dq_0^2\,e^{-q_0^2/M^2}
\left(\begin{array}{c}Im\Lm_l\\ Im\Lm_t\end{array}\right)~.
\ee
Defining new variables $\lm$ and $u$ by $q_0=\lm|\vq|$ and $|\vq|x=u$,
it becomes
\be
\left(\begin{array}{c}\Lm_l\\ \Lm_t\end{array}\right)=
\frac{|\vq|^2}{8(2\pi)^2M^2}\int_0^1\,d\lm^2\,e^{-\lm^2|\vq|^2/M^2}
\lm\int_{|\vq|}^{\infty}du\left(\begin{array}{c}(1-\lm^2)u^2/2\\
-(1-\lm^2)^2|\vq|^4/6\end{array}\right)\cdot D
\ee
where
\bea 
D&=&\frac{1}{\lm|\vq|}\left(\frac{1}{e^{\bet(u-\lm|\vq|)/2}-1}
-\frac{1}{e^{\bet(u+\lm|\vq|)/2}-1}\right) \nonumber \\
 &\rw &-2\frac{d}{du}\left(\frac{1}{e^{\bet u/2}-1}\right),
\eea
as $|\vq|\rw 0$. Thus the short cut also cannot contribute for $\vq=0$.

The $\pi\pi$ loop in the self-energy correction is given by Eq.(3.12) where
$\Pi_{\mn}$ is of the form of Eq.(A.1) with $c=(2G_\rho/\mv\F^2)^2$
and $\fmn=C_{\mn}$. From the previous treatment it is clear that this
amplitude again does not lead to any corrections of $O(T^2)$.

\section*{Acknowledgement}

One of us (S.M.) acknowledges support of CSIR, Government of India.

\end{document}